\documentclass{PoS}

\usepackage{amsmath,graphicx,verbatim,epsfig}
\usepackage{epstopdf}
\usepackage[utf8x]{inputenc}
\usepackage{simplewick}

\newcommand{\eps}{\varepsilon}

\newcommand{\nn}{\nonumber}

\newcommand{\bn}{{\overline n}}

\newcommand{\pslash}{{\not \!p}}
\newcommand{\kslash}{{\not \!k}}

\newcommand{\nb}{\overline n}

\newcommand{\veuv}{\varepsilon_{\rm {UV}}}

\newcommand{\be}{\begin{equation}}
\newcommand{\ee}{\end{equation}}
\newcommand{\bea}{\begin{eqnarray}}
\newcommand{\eea}{\end{eqnarray}}
\newcommand{\balign}{\begin{align}}
\newcommand{\ealign}{\end{align}}
\newcommand{\as}{\alpha_s}

\newcommand{\sandwich}[3]{\left< #1 \right | #2 \left | #3 \right >}

\newcommand{\bg}{\begin{gather}}
\newcommand{\foma}{\end{gather}}

\newcommand{\noopsort}[1]{}

\def\ve{\varepsilon}

\def\pd{\partial}

\def\L{\Lambda}

\def\z{\zeta}

\def\<{\langle}
\def\>{\rangle}

\def\a{\alpha}

\def\g{\gamma}  \def\G{\Gamma}
\def\d{\delta}  \def\D{\Delta}
   \def\L{\Lambda}

\def\m{\mu}

\def\z{\zeta}

\def\({\left(}
\def\[{\left[}
\def\){\right)}
\def\]{\right]}

\def\ln{\hbox{ln}}
\def\log{\hbox{log}}

\def\Qslash{Q\!\!\!\!\!\slash}
\def\Dslash{D\!\!\!\!\!\slash}

\def\nslash{n\!\!\!\!\slash}
\def\bnslash{\overline n\!\!\!\!\slash}
\def\pslash{p\!\!\!\!\slash}
\def\bpslash{\overline p\!\!\!\!\slash}

\def\kslash{k\!\!\!\!\slash}

\def\pdslash{\partial\!\!\!\slash}
\def\Aslash{A\!\!\!\!\slash}

\def \le { \left    }
\def \ri { \right }

\def\bp{\overline p}

\title{Definition and Evolution of Transverse Momentum Distributions}
\ShortTitle{TMDs}

\author{Miguel G. Echevarr\'ia\\
Departamento de F\'isica Te\'orica II,
Universidad Complutense de Madrid (UCM),
28040 Madrid, Spain\\
E-mail: \email{miguel.gechevarria@fis.ucm.es}}
\author{\speaker{Ahmad Idilbi}\\%\thanks{A footnote may follow.}
Instit\"{u}t f\"{u}r Theoretische Physik, Universit\"at Regensburg,
D-93040 Regensburg, Germany\\
E-mail: \email{ahmad.idilbi@physik.uni-regensburg.de}}
\author{\speaker{Ignazio Scimemi}\\
Departamento de F\'isica Te\'orica II,
Universidad Complutense de Madrid (UCM),
28040 Madrid, Spain\\
E-mail: \email{ignazios@fis.ucm.es}}
\abstract{
We consider the definition of unpolarized transverse-momentum-dependent parton distribution functions while staying on-the-light-cone. By imposing a requirement of identical treatment of  two collinear sectors, our approach, compatible with a generic factorization theorem with the soft function included, is valid for all non-ultra-violet regulators (as it should), an issue which causes much confusion in the whole field. We explain how large logarithms can be resummed in a way which can be considered as an alternative to the use of Collins-Soper evolution equation. The evolution properties are also discussed and the gauge-invariance, in both classes of gauges, regular and singular, is emphasized.
}

\FullConference{
QCD Evolution Workshop\\
May 14 - 17, 2012\\
Thomas Jefferson National Accelerator Facility,
Newport News, VA (USA)}

\begin{document}
%%%%%%%%%%%%%%%%%%%%%%%%%%%%%%%%%%%%
%%%%%%%%%%%%%%%%%%%%%%%%%%%%%%%%%%%%
%%%%%%%%%%%%%%%%%%%%%%%%%%%%%%%%%%%%
\section{Introduction}
In high-energy physics, transverse-momentum-dependent parton distribution functions (TMDPDFs), with or without spin dependence, have proved to be an essential quantities for unraveling the internal structure of protons \cite{Barone:2010zz}, as well as being an ingredients representing hadronic physics in a wide class of factorized physical observables. In this effort we will consider the unpolarized TMDPDF and how it should (or could) be defined in a way compatible with a generic factorization theorem of a transverse-momentum-dependent (TMD) physical observable. To properly define such quantities one needs to consider the following issues combined:
\begin{enumerate}
\item The role of the soft function, to be defined below, as a crucial part of the factorized hadronic tensor.
\item The distinction between pure and naively calculated collinear matrix elements.
\item How a definition of a TMDPDF should be independent of the use of any regulator of the non-ultra-violet divergences (nUV).
By definition, ``nUV divergences'' include the physical infra-red (IR) ones of perturbative QCD (pQCD) and the un-physical ones like rapidity divergences.
\end{enumerate}

Consider the Drell-Yan heavy lepton pair production process where the lepton pair is produced with transverse momentum $q_T$ much larger than $\Lambda_{\rm QCD}$. Assuming that the soft and collinear modes completely capture the IR of full QCD, which is a highly nontrivial statement, then, at leading twist, one can establish the following factorization theorem for $q_T$-dependent cross section:
\begin{align}\label{eq:fact0}
\sigma=H(Q^2/\mu^2) J_n\, J_{\bn}\, S \,\,,
\end{align}
where the $q_T$-dependence is implicit in the pure collinear matrix elements $J_{n,\bn}$ and the soft function $S$. By ``pure'' collinear contribution we mean that the soft (zero-bin) contamination has to be subtracted out \cite{Manohar:2006nz}. In the remainder of this section we refer to all quantities in Eq.~(\ref{eq:fact0}) as the partonic version of the physical matrix elements unless otherwise stated. To explain in simple terms how the TMDPDF should be defined we emphasize the following:

1. In pQCD the extraction of the hard part is performed by considering a generic Feynman diagram contributing to $\sigma$ and then ``subtracting'' all the physics contributing to the relevant lower scales. In the effective field theory (EFT) methodology this is equivalent to a multiple matching procedure starting from full QCD and the appropriate effective theories. In both approaches it is obvious that the hard part should depend only on the hard scale and the renormalization one as well. Thus, $H$ is a polynomial  of only {\emph {one}} quantity: $\log(Q^2/\mu^2)$. Any dependence on nUV regulators needed to calculate $\sigma$ is prohibited.

2. The un-canceled nUV divergences in $\sigma$, no matter how they are regulated, are completely a genuine IR ones that has to be generated by similar ones from the right-hand side of Eq.~(\ref{eq:fact0}). This is the ultimate check (at least pertubatively) for either the subtraction philosophy or the EFT matching procedure to work! The immediate conclusion is that the product $J_n\,J_{\bn}\,S$ has to include only the IR of QCD. Again we emphasize that this simple observation is (or should be, if perturbative calculations are performed properly) independent of any regulator(s) of the nUV divergences.

Given the factorization theorem of $\sigma$, where $\sigma$ is calculated in full QCD, with the above two observations and combined with the requirement of identical treatment among the two collinear sectors $n$ and $\bn$, one can easily conclude that each quantity, $J_n\,\sqrt{S}$ and $J_{\bn}\,\sqrt{S}$, has to be a well-defined quantity in the sense that it is free of any nUV divergences that do not exist in the full pQCD calculation of $\sigma$ (a calculation that is needed in order to extract the hard part $H$). Again we mention that this line of reasoning is valid no matter which set of regulators is used for the nUV as long as one uses the same set for both collinear sectors. Thus we define the quantity $F_{n(\bn)}=J_{n,(\bn)}\,\sqrt{S}$ as the ``TMDPDF''. For two identical partons in the initial states one simply has: $F_n=F_{\bn}$

Notice that no attempt to go ``off-the-light-cone'' has been made. There are mainly two reasons for that. First is that one is completely free to regularize the nUV divergences with \emph{any} set of regulators and moreover there is also no need to even distinguish between the different kinds of divergences in that set. This is definitely true if one believes that the factorized result of $\sigma$ holds to all orders in perturbation theory. The second reason is that large logarithms can be resummed (and one should be able to resum them) without any reference to a specific set of regulators. We believe this observation is simple enough and it should be undisputed.

We finally remark that if one insists on introducing different regulators for the two collinear directions (like different off-shellnesses, $\Delta$'s, etc.), as it is done in~\cite{Chay:2012mh}, then \emph{certain} logarithms of the ratio of those different {$n$,$\bn$}-regulators will still appear in each one of the TMDPDFs, $F_n$ and $F_{\bn}$. Those logarithms are intimately related to the existence of the  Wilson lines in the collinear and soft matrix elements. However those logarithms will cancel when one combines the two collinear contributions and the soft function to $\sigma$. Thus we find it completely harmless to use identical parameters as long as one can still resum large logarithms and extract each TMDPDF from the relevant experimental data.

We start by briefly reviewing Collins definition of the TMDPDF \cite{Collins:2011zzd} and then compare it with the ``on-the-light-cone'' one \cite{GarciaEchevarria:2011rb}.
Some phenomenological applications of the  definition and evolution of TMDs as defined in \cite{GarciaEchevarria:2011rb} can be found in Ref.~\cite{Echevarria:2012pw}.

%%%%%%%%%%%%%%%%%%%%%%%%%%%%%%%%
%%%%%%%%%%%%%%%%%%%%%%%%%%%%%%%%
%%%%%%%%%%%%%%%%%%%%%%%%%%%%%%%%%
\section{Collins Approach}
%%%%%%%%%%%%%%%%%%%%%%%%%%%%%%%%%

In impact parameter space, Collins defines the TMDPDF as:
\begin{align}
\label{cc}
\tilde F_n(x,b;Q,y_n) &=
\tilde {\hat J}_n(x,b;Q^2)
\sqrt{\frac{\tilde S(b;+\infty,y_n)}{\tilde S(b;+\infty,-\infty)\, \tilde S(b;y_n,-\infty)}}
=
\stackrel{\rm on-the-LC}{\overbrace{
\frac{\tilde {\hat J}_n(x,b;Q^2)}{\sqrt{\tilde S(b;+\infty,-\infty)}}
}}
\stackrel{\rm off-the-LC}{\overbrace{
\sqrt{\frac{\tilde S(b;+\infty,y_n)}{\tilde S(b;y_n,-\infty)}}
}}
\,.
\end{align}
The first factor in the last term is exactly the definition of the TMDPDF where all Wilson lines are still on-the-light-cone. However it should be noted that this definition relies on the hypothesis that soft contamination in the naively calculated collinear contribution $\tilde {\hat J}_n$ can be removed by soft function subtraction. This issue is a subtle one and its validity depends on the consistent use of the nUV regulators of the naive collinear, its soft limit (or its zero-bin) and the soft function built from soft Wilson lines. More discussion on this can be found in \cite{Lee:2006nr,Idilbi:2007ff,Idilbi:2007yi}. Hereafter we will not discuss this issue further.
The aim of the second term, is \emph{not}  canceling  the rapidity divergencies as one might naively think (especially when going off-the-light-cone) but to introduce a new parameter $y_n$ that allows the derivation of a ``Collins-Soper'' like evolution equation to resum large logarithms of $Q/k_T$. The dependence (of Collins' TMDPDF) on this arbitrary parameter will cancel only when two TMDPDFs (or, more generally, two ``collinear'' hadronic contributions like one TMDPDF and one fragmentation function in the case of SIDIS) are combined since the two sectors are tilted asymmetrically (notice the $e^{\pm 2y_n}$ in the two tilted collinear directions) so as to achieve this cancelation. We want now to raise two important points:
\begin{enumerate}
\item The second factor in the last term of Eq.~(\ref{cc}) introduces an arbitrary dependence on $y_n$ of a single TMDPDF. In all existing literature using Collins formalism (e.g.,\cite{Aybat:2011zv,Aybat:2011ge,Aybat:2011ta,Anselmino:2012aa}) for single sector, this dependence is canceled when $y_n$ is set to zero, which is to say that we are back on-the-light-cone since that factor reduces to unity.
\item If that factor is exactly $1$ on-the-light-cone, and one can still resum large logarithms by simply noticing that the first factor in the last term of Eq.~(\ref{cc}) is free from all un-physical nUV divergences (under the requirement of identical treatment of two collinear sectors) then why one needs to consider the $y_n$-dependent factor in the first place?
Life is much simpler without it!
\end{enumerate}

Given all the considerations above, we now show more explicitly how Collins approach reduces to ours when one chooses $y_n=0$. In that approach the master equation for quark TMDPDF, denoted by $\tilde{F}_{f/P}^{C}$, is
\begin{multline}
\label{eq:TMDcollins}
\tilde{F}_{f/P}^{C}(x,b;\zeta_{F,f},\m_f=Q_f) =
%\stackrel{\rm A}{\overbrace{
\sum_{j=q,g} \int_x^1 \frac{d x'}{x'}
\tilde{C}_{f/j}^{C}\le(\frac{x}{x'},b;\z_{F,i}=\m_{I}^2,\mu_{I}\ri)\,
\phi_{j/P}(x';\mu_{I})
%}}
\nn\\
\times
%\stackrel{\rm B}{\overbrace{
\exp \left\{
\tilde{K}(b;\mu_{I})\, \ln \frac{\sqrt{\zeta_{F,f}}}{\mu_{I}}  +
\int_{\mu_{I}}^{\m_f=Q_f} \frac{d \mu'}{\mu'} \left[ \g_F^{C}\le(\as(\mu');\ln\frac{\z_{F,f}}{\m'^2}\ri) \right]\right\}
%}}
%\times
%\stackrel{\rm C}{\overbrace{
%\tilde{F}_{f/P}^{NP}(x,b;\z_{F,f},\z_{F,0})
%}}
\,.
\end{multline}
where the matching of the quark TMDPDF onto the standard PDF is
\begin{align}
\tilde F_{f/P}(x,b;\z_F,\m) &=
\sum_{j=q,g}
\int_x^1 \frac{dx'}{x'}\, \tilde C_{f/j}^{C}(\frac{x}{x'},b;\z_F,\m)\,
\phi_{j/P}(x';\m)
+ {\cal O}\le((\L_{QCD}b)^a\ri)\,.
\end{align}
On the other hand, our master equation for the TMDPDF on-the-light-cone is
\begin{multline}
\label{eq:TMDours}
\tilde{F}_{f/P}(x,b;Q_f^2,\m_f=Q_f) =
%\stackrel{\rm A}{\overbrace{
\sum_{j=q,g} \int_x^1 \frac{d x'}{x'}
\tilde{C}_{f/j}^{\Qslash}\le(\frac{x}{x'},b;\m_{I}\ri)\,
\phi_{j/P}(x';\mu_{I})
%}}
\nn\\
\times
%\stackrel{\rm B}{\overbrace{
\exp \left\{
-D(b;\mu_{I})\, \ln \frac{Q_f^2}{\m_{I}^2}  +
\int_{\mu_{I}}^{\m_f=Q_f} \frac{d \mu'}{\mu'} \left[ \g_F\le(\as(\mu');\ln\frac{Q_f^2}{\m'^2}\ri) \right]\right\}
%}}
%\times
%\stackrel{\rm C}{\overbrace{
%\tilde{F}_{f/P}^{NP}(x,b;\z_{F,f},\z_{F,0})
%}}
\,.
\end{multline}
where the matching of the quark TMDPDF onto the standard PDF is given by
\begin{align}
\tilde F_{f/P}(x,b;Q_f^2,\m) &=
\sum_{j=q,g}
\int_x^1 \frac{dx'}{x'}\, \tilde C_{f/j}(\frac{x}{x'},b;Q_f^2,\m)\,
\phi_{j/P}(x';\m)
+ {\cal O}\le((\L_{QCD}b)^a\ri)\,.
\end{align}
The fact that the TMDPDF $\tilde F_{f/P}$ is free from all un-physical nUV divergencies allows us to write
\begin{align}
\tilde C_{f/j}(x,b;Q^2,\m) &=
\le(\frac{Q^2 b^2}{4e^{-2\g_E}} \ri)^{-D(b;\m)} \tilde C_{f/j}^{\Qslash}(x,b;\m)\,.
\end{align}

In both cases $\m_I$ should be chosen so that the large logarithms in the matching coefficients $\tilde C$ are canceled. This choice can follow the well-known CSS approach~\cite{Collins:1984kg}, where one relates $\m$ with $b$ and has to deal with the issue of Landau pole. However one can also follow the work of \cite{Becher:2010tm} and transform back to momentum space before setting the scale $\m_I$, which is related finally to $k_T$.

When setting $y_n=0$  one gets the following relations:
\begin{align}
\tilde K(b;\m) &\longrightarrow -2D(b;\m)\,,\quad\quad
\g_F^{C}(\z,\m) \longrightarrow \g_F(Q^2,\m)\,,\quad\quad
\tilde C_{f/j}^{C}(x,b;\z,\m) \longrightarrow \tilde C_{f/j}(x,b;Q^2,\m)\,.
\end{align}
The above relations hold to all orders in perturbation theory. Thus we see that Collins formalism reduces to the ``on-the-light-cone'' one.

%%%%%%%%%%%%%%%%%%%%%%%%%%%%%%%%
%%%%%%%%%%%%%%%%%%%%%%%%%%%%%%%%
\section{The TMDPDF On-The-Light-Cone}
\label{sec:TMDPDF}
%%%%%%%%%%%%%%%%%%%%%%%%%%%%%%%%

For many years, the two main issues that people have tried to resolve in order to get the correct definition of the TMDPDF are the following:
\begin{itemize}
\item On one hand, the TMDPDF should be free from rapidity divergencies. In the case of the PDF, as it is shown in Sec.~\ref{sec:pdf}, these divergencies cancel when one combines virtual and real diagrams. But when we allow the collinear matrix element to depend also on the transverse coordinate, such cancelation does not hold anymore.
\item On the other hand, the TMDPDF is a quantity that depends on two different scales: the hard probe $Q$ and the transverse momentum $k_T$. When these scales are widely separated, one needs to resum the large logarithms appearing in perturbative expression of the TMDPDF.
\end{itemize}

In the following we give a consistent definition of the TMDPDF, while staying on-the-light-cone that addresses the issues mentioned above. More details can be found in \cite{GarciaEchevarria:2011rb}.

An intermediate step towards getting the factorization theorem of the hadronic tensor $M$ is given by
\begin{align}\label{eq:fact1}
M &=
H(Q^2/\mu^2)\,
\int\!d^4y\,e^{-iq\cdot y}\,
J_n(0^+,y^-,\vec y_\perp)\,
J_\bn(y^+,0^-,\vec y_\perp)\,
S(0^+,0^-,\vec y_\perp)\,,
\end{align}
where $H$ is the hard matching coefficient, $J_{n(\bn)}$ are the pure-collinear matrix elements and $S$ is the soft function.
The collinear pieces $J_{n(\bn)}$ are intended to contain just collinear modes, however in actual calculation we integrate over all momentum space thus collecting a contribution from soft modes as well. Below we denote the ``naive collinear'' as: $\hat J_{n(\bn)}$.

The fact that the soft function depends just on the transverse coordinate was derived in~\cite{GarciaEchevarria:2011rb} based on power counting arguments of the soft and collinear modes.

Based on Eq.~(\ref{eq:fact1}), by symmetry, we define the TMDPDF as
\begin{align}\label{eq:tmddef}
F_n(x;\vec k_{n\perp}) &=\frac{1}{2}
\int \frac{dr^-d^2\vec r_\perp}{(2\pi)^3} e^{-i(\frac{1}{2}r^-xp^+-\vec r_\perp \cdot \vec k_{n\perp})}
J_n(0^+,r^-,\vec r_\perp)\sqrt{S(0^+,0^-,\vec r_\perp)}
\nn\\
&=
\int \frac{dr^-d^2\vec r_\perp}{(2\pi)^3} e^{-i(\frac{1}{2}r^-xp^+-\vec r_\perp \cdot \vec k_{n\perp})}
\frac{\hat J_n(0^+,r^-,\vec r_\perp)}{\sqrt{S(0^+,0^-,\vec r_\perp)}}\,,
\end{align}
where the second equality holds only if the subtraction of soft contamination is equivalent to dividing by the soft function.

We can use any regulator to regularize the nUV, however one has to be consistent. A factorization theorem can be understood from the effective field theory point of view as a multiple steps of matching between different effective theories. In this sense it is a must that one regularizes the ``IR'' consistently in the theory above and below the relevant scale. Then, the matching coefficients ($H$ and $\tilde C$ in our case) can never depend on the set of the nUV regulators.

We choose a frame where $p=(p^+,0^-,0_\perp)$ and $\bp=(0^+,\bp^-,0_\perp)$ with $p^+=\bp^-=Q$, and write the poles of fermion propagators with a real and positive parameter $\D$:
\begin{align}\label{fermionsDelta}
\frac{i(\pslash+\kslash)}{(p+k)^2+i0} \longrightarrow
\frac{i(\pslash+\kslash)}{(p+k)^2+i\D}\,,
\quad\quad
\frac{i(\bpslash+\kslash)}{(\bp+k)^2+i0} \longrightarrow
\frac{i(\bpslash+\kslash)}{(\bp+k)^2+i\D}\,.
\end{align}
The above prescription applies as well to the fermion propagators. The corresponding pole-shifting for collinear and soft Wilson lines goes as follows
\begin{align}
\frac{1}{k^+\pm i0} \longrightarrow
\frac{1}{k^+\pm i\d}\,,
\quad\quad
\frac{1}{k^-\pm i0} \longrightarrow
\frac{1}{k^-\pm i\d}\,,
\end{align}
where $\d$ is related to $\D$ through the large components of the collinear fields,
\begin{align} \label{regul_DeltaDY}
\d = \frac{\D}{p^+} = \frac{\D}{\bp^-}\,.
\end{align}
This relation comes from the fact that in order to recover the correct IR structure of QCD, the contributions of collinear and soft Wilson lines have to be consistent with the collinear and soft limits, respectively, of full QCD.

%%%%%%%%%%%%%%%%%%%%%%%%%%%%%%%%%
\subsection{Virtual Diagrams}
\label{sec:virtual}
%%%%%%%%%%%%%%%%%%%%%%%%%%%%%%%%%
%%%%%%%%%%%%%%%%%%%%%%%%%%%%%FIGURE
\begin{figure}
\begin{center}
\includegraphics[width=0.8\textwidth]{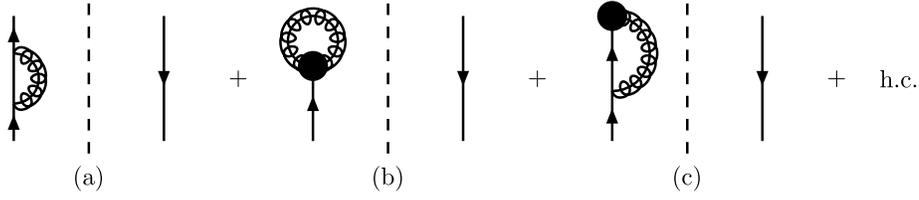}
\end{center}
\caption{\it Virtual corrections for the collinear matrix element. The black blobs represent the collinear
Wilson lines $W$ in Feynman gauge or the $T$ Wilson lines in light-cone gauge. Curly propagators with a line stand for collinear gluons. ``h.c.'' stands for Hermitian conjugate.}
\label{n_virtuals}
\end{figure}
%%%%%%%%%%%%%%%%%%%%%%%%%%%%%%%%%ENDFIGURE
%%%%%%%%%%%%%%%%%%%%%%%%%%%%%%%%%%%%%%%%%%%%FIGURE
\begin{figure}
\begin{center}
\includegraphics[width=0.7\textwidth]{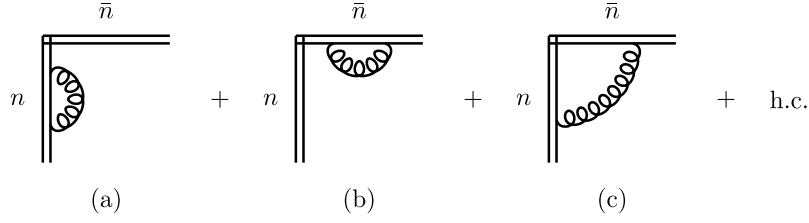}
\end{center}
\caption{\it Virtual corrections for the soft function. Double lines represent the soft Wilson lines, $S_{n(\bn)}$. ``h.c.'' stands for  Hermitian conjugate.}
\label{s_virtuals}
\end{figure}
%%%%%%%%%%%%%%%%%%%%%%%%%%%%%%%%%ENDFIGURE

The diagrams in figs.~(\ref{n_virtuals}) and~(\ref{s_virtuals}) give collinear  and soft  virtual contributions respectively to $F_n$. The Wave Function Renormalization (WFR) diagram~(\ref{n_virtuals}a) and its Hermitian conjugate give
\begin{align}\label{1a}
\hat J_{n1}^{(\ref{n_virtuals}a)+(\ref{n_virtuals}a)^*}&=
\frac{\alpha_s C_F}{2\pi}
\d(1-x)\d^{(2)}(\vec k_{n\perp})
\le[ \frac{1}{\veuv} + \ln\frac{\m^2}{\D}+\frac{1}{2} \ri]
\end{align}
The $W$ Wilson line tadpole diagram, (\ref{n_virtuals}b), is identically $0$, since $\bn^2=0$.
Diagram~(\ref{n_virtuals}c) and its Hermitian conjugate give
\begin{align}
\label{eq:jn1c}
\hat J_{n1}^{(\ref{n_virtuals}c)+(\ref{n_virtuals}c)^*}&=
%-2ig^2C_F \d(1-x)\d^{(2)}(\vec k_{n\perp}) \m^{2\e} \int \frac{d^dk}{(2\pi)^d}
%\frac{p^++k^+}{[k^+-i\d^+][(p+k)^2+i\D^-][k^2+i0]}
%+ h.c.\nn\\
%&=
\frac{\a_s C_F}{2\pi}
\d(1-x)\d^{(2)}(\vec k_{n\perp})
\left[
\frac{2}{\veuv}\ln\frac{\d}{p^+} + \frac{2}{\veuv} - \ln^2\frac{\d\D}{p^+\m^2}
- 2\ln\frac{\D}{\m^2} + \ln^2\frac{\D}{\m^2} + 2 - \frac{7\pi^2}{12}
\right]\, .
\end{align}
The contribution of diagrams~(\ref{s_virtuals}a) and~(\ref{s_virtuals}b) is zero, since~(\ref{s_virtuals}a) is proportional to $n^2=0$
and~(\ref{s_virtuals}b) to $\bn^2=0$.
The diagram~(\ref{s_virtuals}c) and its Hermitian conjugate give
\begin{align}\label{s_virtuals_c}
S_1^{(\ref{s_virtuals}c)+(\ref{s_virtuals}c)^*}&=
%-2ig^2 C_F \d^{(2)}(\vec k_{n\perp}) \mu^{2 \eps}
%\int \frac{d^d k}{(2 \pi)^d} \frac{1}{[k^+-i\d^+] [k^-+i\d^-] [k^2+i0]} +h.c.
%\nn \\
%&=
- \frac{\alpha_s C_F}{2\pi}
\d^{(2)}(\vec k_{n\perp})
\left[\frac{2}{\veuv^2}-\frac{2}{\veuv}\ln\frac{\d^2}{\mu^2}+
\ln^2\frac{\d^2}{\mu^2}+\frac{\pi^2}{2}\right]\, .
\end{align}

The virtual part of the TMDPDF at ${\cal O}(\as)$ while using the relation in Eq.~(\ref{regul_DeltaDY}) is
\begin{align}
\label{eq:jvn}
 F_{n1}^v &= - \frac{1}{2} \hat J_{n1}^{(\ref{n_virtuals}a)+(\ref{n_virtuals}a)^*} + \hat J_{n1}^{(\ref{n_virtuals}c)+(\ref{n_virtuals}c)^*}
 - \frac{1}{2} \d(1-x) S_1^{(\ref{s_virtuals}c)+(\ref{s_virtuals}c)^*}
\nn\\
&=
\frac{\alpha_s C_F}{2 \pi}
\d(1-x)\d^{(2)}(\vec k_{n\perp})
\left[
\frac{1}{\veuv^2} + \frac{1}{\veuv} \left( \frac{3}{2} + \ln\frac{\mu^2}{Q^2} \right)
%\right.
%\nn\\
%&\left.
-\frac{3}{2}\ln\frac{\D}{\mu^2} - \frac{1}{2}\ln^2\frac{\D^2}{Q^2\m^2} + \ln^2\frac{\D}{\m^2}
+ \frac{7}{4} - \frac{\pi^2}{3}\right]\,.
\end{align}
As mentioned earlier, individual contributions to $F_{n1(\bn 1)}^v$ have mixed divergences, however $F^v_{n1}$ itself is free from them. The $\D$-dependence that remains is pure IR, as can be seen from the matching with full QCD in Sec.~\ref{sec:hard}.

%%%%%%%%%%%%%%%%%%%%%%%%%%%%%%%%%
\subsection{Real Diagrams}
\label{sec:real}
%%%%%%%%%%%%%%%%%%%%%%%%%%%%%%%%%

%%%%%%%%%%%%%%%%%%%%%%%%%%%%%%%%%%%%FIGURE
\begin{figure}
\begin{center}
\includegraphics[width=\textwidth]{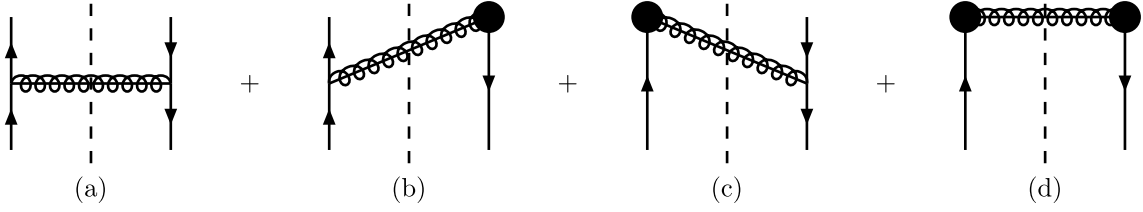}
\end{center}
\caption{\it Real gluon contributions for $\hat J_{n(\bn)}$.
\label{n_reals}}
\end{figure}
%%%%%%%%%%%%%%%%%%%%%%%%%%%%%%%%%%%%ENDFIGURE

\begin{figure}
\begin{center}
\includegraphics[width=\textwidth]{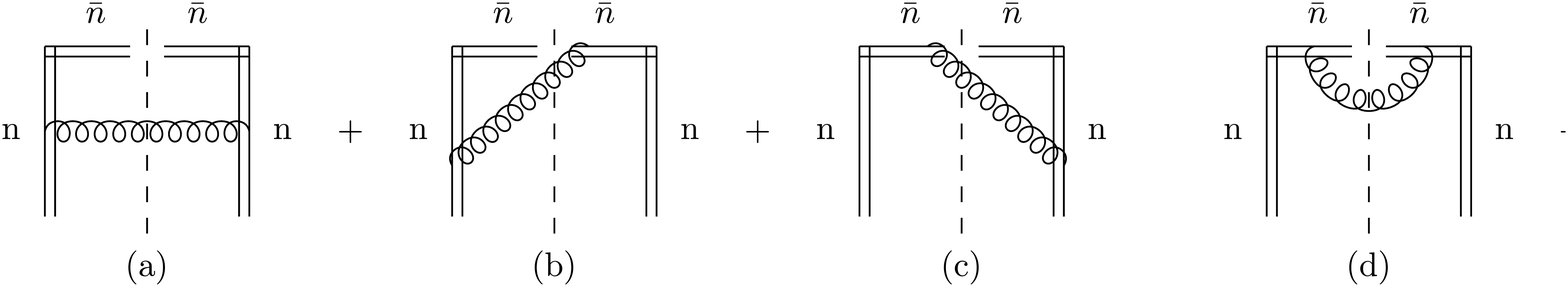}
\end{center}
\caption{\it Real gluon contributions for the Soft function.
\label{s_reals}}
\end{figure}

The relevant diagrams  for the real part of $F_n$ are shown in  figs.~(\ref{n_reals}) and~(\ref{s_reals}). Diagram~(\ref{n_reals}a) gives
\begin{align}\label{eq:jn3a}
\hat J_{n1}^{(\ref{n_reals}a)}&=
%2\pi g^2 C_F p^+ \int\frac{d^dk}{(2\pi)^d}
%\d(k^2)\theta(k^+)\frac{2(1-\ve)|\vec k_\perp|^2}{[(p-k)^2+i\D^-] [(p-k)^2-i\D^-]}
%\nn\\
%&\times
%\d\le((1-x)p^+-k^+\ri) \d^{(2)}(\vec k_\perp+\vec k_{n\perp})
%\nn\\
%&=
\frac{\a_s C_F}{2\pi^2}
(1-\ve)(1-x)
\frac{k_{nT}^2}{\left|k_{nT}^2-i\D(1-x)\right|^2}\,,
\end{align}
The sum of diagram~(\ref{n_reals}b) and its Hermitian conjugate~(\ref{n_reals}c) is
\begin{align}\label{eq:jn3b3c}
\hat J_{n1}^{(\ref{n_reals}b+\ref{n_reals}c)}&=
%-4\pi g^2 C_F p^+  \int\frac{d^dk}{(2\pi)^d}
%\d(k^2)\theta(k^+)\frac{p^+-k^+}{[k^++i\d^+][(p-k)^2+i\D^-]}
%\nn\\
%&\times
%\d\le((1-x)p^+-k^+\ri) \d^{(2)}(\vec k_\perp+\vec k_{n\perp}) + h.c.
%\nn\\
%&=
\frac{\alpha_s C_F}{2\pi^2}
\left[\frac{x}{(1-x)+i\d/p^+}\right]
\left[
\frac{1}{k_{nT}^2-i\D(1-x)}
\right]
+ h.c.
%\nn\\
%&=
%\frac{2\alpha_s C_F}{(2\pi)^{2-2\ve}} \frac{1}{|\vec k_{n\perp}|^2}
%\left[\frac{2x}{(1-x)_+} - 2\delta(1-x) \ln\frac{\d^+}{p^+}\right]
\,,
\end{align}
Diagram~(\ref{n_reals}d) is zero, since it is proportional to $\bn^2=0$.

%%%%%%%%%%%%%%%%%%%%%%%%%%%%%%%%%%%%FIGURE
%%%%%%%%%%%%%%%%%%%%%%%%%%%%%%%%%%%%ENDFIGURE
For the real emission of soft gluons,  diagrams~(\ref{s_reals}a) and~(\ref{s_reals}d) are zero, since they are proportional to $n^2=0$ and $\bn^2=0$ respectively.
Diagram~(\ref{s_reals}b) and its Hermitian conjugate~(\ref{s_reals}c) give
\begin{align}\label{eq:jn4b4c}
S_1^{(\ref{s_reals}b+\ref{s_reals}c)}&=
%-4\pi g^2 C_F  \int\frac{d^dk}{(2\pi)^d}
%\d^{(2)}(\vec k_\perp+\vec k_{n\perp}) \d(k^2)\theta(k^+) \frac{1}{[k^++i\d^+][-k^-+i\d^-]} + h.c.
%\nn\\
%&=
-\frac{\a_s C_F}{\pi^2}
\frac{1}{k_{nT}^2-\d^2} \ln\frac{\d^2}{k_{nT}^2}\,.
%=
%-\frac{4\a_s C_F}{(2\pi)^{2-2\ve}}
%\frac{1}{|\vec k_{n\perp}|^2} \ln\frac{\d^+\d^-}{|\vec k_{n\perp}|^2}\,.
\end{align}

Combining the above contributions according to Eq.~(\ref{eq:tmddef}), one gets the real part of the TMDPDF $F^{r}_{n1}$ at ${\cal O}(\as)$, which is free from rapidity divergencies. A detailed discussion about the remaining pure IR $\D$-dependence can be found in~\cite{GarciaEchevarria:2011rb}.

%%%%%%%%%%%%%%%%%%%%%%%%%%%%%%%%
%%%%%%%%%%%%%%%%%%%%%%%%%%%%%%%%
\section{Integrated PDF}
\label{sec:pdf}
%%%%%%%%%%%%%%%%%%%%%%%%%%%%%%%%
In this section we briefly report the calculation of the integrated PDF at first order in $\as$ with the $\D$-regulator. As it is shown below, the mixed divergencies will cancel when we combine virtual and real contributions, leading to the well-known IR collinear divergence of the PDF.

The virtual diagrams for the PDF are the same as for the naive collinear matrix element that enters into the definition of the TMDPDF, Fig.~(\ref{n_virtuals}). Then we have
\begin{align}
\label{eq:jvn}
\phi_{n1}^v &=
\frac{\alpha_s C_F}{2 \pi}
\d(1-x)
\left[
\frac{1}{\veuv^2} + \frac{1}{\veuv} \left( \frac{3}{2} + \ln\frac{\mu^2}{Q^2} \right)
%\right.
%\nn\\
%&\left.
-\frac{3}{2}\ln\frac{\D}{\mu^2} - \frac{1}{2}\ln^2\frac{\D^2}{Q^2\m^2} + \ln^2\frac{\D}{\m^2}
+ \frac{7}{4} - \frac{\pi^2}{3}\right]\,.
\end{align}

The real diagrams are the same as in Fig.~(\ref{n_reals}), from which we get
\begin{align}
\phi_{n1}^{(\ref{n_reals}a)}&=
%2\pi g^2 C_F p^+ \int\frac{d^dk}{(2\pi)^d}
%\d(k^2)\theta(k^+)\frac{2(1-\ve)|\vec k_\perp|^2}{[(p-k)^2+i\D^-][(p-k)^2-i\D^-]}
%\d\le((1-x)p^+-k^+\ri)
%\nn\\
%&=
\frac{\as C_F}{2\pi} (1-x) \left[ \frac{1}{\veuv} + \ln\frac{\m^2}{\D} - 1 - \ln(1-x)\right]\,,
\end{align}
and
\begin{align}
\phi_{n1}^{(\ref{n_reals}b+\ref{n_reals}c)}&=
%-4\pi g^2 C_F p^+  \m^{2\ve}\int\frac{d^dk}{(2\pi)^d}
%\d(k^2)\theta(k^+)\frac{p^+-k^+}{[k^++i\d^+][(p-k)^2+i\D^-]}
%\nn\\
%&\times
%\d\le((1-x)p^+-k^+\ri) + h.c.
%\nn \\
%&=
\frac{\as C_F}{2\pi} \left[
\left( \frac{1}{\veuv} + \ln\frac{\m^2}{\D} \right)
\left( \frac{2x}{(1-x)_+} - 2\d(1-x)\ln\frac{\D}{Q^2} \right)
\right.
\nn\\
&\left.
- 2\d(1-x)\left(1-\frac{\pi^2}{24}-\frac{1}{2}\ln^2\frac{\D}{Q^2}\right)
+ \frac{\pi^2}{2} \d(1-x)
\right]\,,
\end{align}
Diagram~(\ref{n_reals}d) is zero since it is proportional to $\bn^2$.

Mixed divergencies $\frac{1}{\veuv}\ln\frac{\D}{Q^2}$ (and also double IR poles $\ln^2\frac{\D}{Q^2}$) are cancelled when we combine virtual and real contributions. The PDF to first order in $\as$ is finally
\begin{align}
\label{eq:pdfdelta}
\phi_{n1}(x;\m) &= \delta(1-x)+\frac{\as C_F}{2\pi} \left[
{\cal P}_{q/q} \left( \frac{1}{\veuv} - \ln\frac{\D}{\m^2}\right)
- \frac{1}{4}\d(1-x) - (1-x)\left[1+\ln(1-x)\right]
\right]\,,
\end{align}
where ${\cal P}_{q/q}$ is the one-loop quark splitting function of a quark in a quark,
\begin{align}
{\cal P}_{q/q} = \left( \frac{1+x^2}{1-x} \right)_+ =
\frac{1+x^2}{(1-x)_+} + \frac{3}{2}\d(1-x) =
\frac{2x}{(1-x)_+} + (1-x) + \frac{3}{2}\d(1-x)\,.
\end{align}
The IR collinear divergence is encoded in the single logarithm $\ln(\D/\m^2)$, while the rest is a function of $x$ that in general depends on the regulator used.

%%%%%%%%%%%%%%%%%%%%%%%%%%%%%%%%
\section{$Q^2$-Dependence and Resummation}
\label{sec:q2}
%%%%%%%%%%%%%%%%%%%%%%%%%%%%%%%%
When $k_T \gg \L_{QCD}$ we can perform an operator product expansion (OPE) of the TMDPDF onto the integrated PDF,
\begin{align}\label{eq:ope}
\tilde F_n(x;b,Q,\m) = \int_x^1 \frac{dx'}{x'} \tilde C_n\le(\frac{x}{x'};b,Q,\m\ri)\, \phi_n(x';\m)\,,
\end{align}
where
\begin{align}
\phi_n(x;\m) = \frac{1}{2} \int \frac{dy^-}{2\pi} e^{-i\frac{1}{2}y^-xp^+}
\sandwich{p}{\bar \chi_n(0^+,y^-,\vec 0_\perp) \frac{\bnslash}{2}\chi_n^\dagger(0^+,0^-,\vec 0_\perp)}{p}
|_\textrm{zb~ included}\,,
\end{align}
and $\tilde C_n$ is the matching coefficient that cannot depend on any IR regulator as mentioned before. In this section we compute $\tilde C_n$ to first order in $\as$.

The virtual part of the TMDPDF in momentum space was given in Eq.~(\ref{eq:jvn}), and in impact parameter space it reads
\begin{align}
\tilde F_{n1}^v &=
\frac{\alpha_s C_F}{2 \pi}
\d(1-x)
\left[
\frac{1}{\veuv^2} + \frac{1}{\veuv} \left( \frac{3}{2} + \ln\frac{\mu^2}{Q^2} \right)
- \frac{3}{2}\ln\frac{\D}{\mu^2} - \frac{1}{2}\ln^2\frac{\D^2}{Q^2\m^2}
+ \ln^2\frac{\D}{\m^2} + \frac{7}{4}-\frac{\pi^2}{3}\right]\,.
\end{align}
The Fourier transform of real-gluon emission diagrams given in Eqs.~(\ref{eq:jn3a}, \ref{eq:jn3b3c}, \ref{eq:jn4b4c}), while keeping the $\D$'s to regulate the nUV divergences, are
\begin{align}
\tilde{\hat J}_{n1}^{(\ref{n_reals}a)}&=
\frac{\as C_F}{2\pi}(1-x)\, \ln\frac{4e^{-2\g_E}}{\D(1-x) b^2}\,,
\end{align}
\begin{align}
\tilde{\hat J}_{n1}^{(\ref{n_reals}b+\ref{n_reals}c)}&=
\frac{\as C_F}{2\pi} \left[
\ln\frac{4e^{-2\g_E}}{\D b^2}
\left( \frac{2x}{(1-x)_+}-2\d(1-x)\ln\frac{\D}{Q^2} \right)
+\frac{\pi^2}{2}\d(1-x)
\right.
\nn\\
&\left.
-2\d(1-x) \left(
1 - \frac{\pi^2}{24} - \frac{1}{2}\ln^2\frac{\D}{Q^2}
\right)
\right]\,,
\end{align}
and
\begin{align}
\tilde S_1^{(\ref{s_reals}b+\ref{s_reals}c)}&=
\frac{\as C_F}{2\pi}
\left( \ln^2\frac{4e^{-2\g_E}Q^2}{\D^2 b^2} + \frac{2\pi^2}{3} \right)\,.
\end{align}

Finally, the TMDPDF in impact parameter space to first order in $\as$ is
\begin{align}
\tilde F_{n}
&=\phi_n+\frac{\alpha_s C_F}{2\pi}\Big[ -L_T{\cal P}_{q/q}+(1-x) -\delta(1-x)
\left(\frac{1}{2}L_T^2-\frac{3}{2}L_T+L_T\ln\frac{Q^2}{\mu^2}+\frac{\pi^2}{12}\right)\Big]\,,
\end{align}
where $L_T = \ln(\m^2b^2e^{2\g_E}/4)$, $\phi_n$ is the PDF given in Eq.~(\ref{eq:pdfdelta}) and the remaining part is exactly the OPE matching coefficient
\begin{align}\label{coeff}
\tilde C_n(x;b,Q,\m) &=  \d(1-x) + \frac{\alpha_s C_F}{2 \pi}
\left[-{\cal P}_{q/q} L_T + (1-x)
- \d(1-x)\left( \frac{1}{2}L_T^2 - \frac{3}{2} L_T + \ln\frac{Q^2}{\m^2}L_T
+\frac{\pi^2}{12}  \right) \right]\,.
\end{align}

In the integrated PDF the mixed divergencies are canceled when we combine virtual and real diagrams. When we allow for a transverse separation also in the collinear matrix element, real diagrams do not cancel these divergencies of the virtual part. It is the soft function (square root) that does this job. Mixed divergencies in the virtual part of the collinear matrix element are canceled by the virtual part of the soft function; and the ones in the real part of the collinear are canceled by the real part of the soft function. All the remaining $\D$-dependence is pure IR.

The fact that the TMDPDF is free from rapidity divergencies allows us to extract and exponentiate its $Q^2$-dependence to all orders in the following way~\cite{GarciaEchevarria:2011rb},
\begin{align}\label{eq:q2factor}
\tilde F_n(x;\vec b_\perp,Q,\m) &= \left( \frac{Q^2 b^2}{4e^{-2\g_E}} \right)^{-D(\as,L_T)} \tilde C^{\Qslash}_n(x;\vec b_\perp,\m)
\otimes \phi_n(x;\m)\,,
\end{align}
where
\begin{align}
\tilde C^{\Qslash}_n(x;\vec b_\perp,\m)& =
\delta(1-x)+\frac{\alpha_s C_F}{2 \pi}\left[-{\cal P}_{q/q} L_T + (1-x) - \delta(1-x)
\left(-\frac{1}{2}L_T^2-\frac{3}{2}L_T+\frac{\pi^2}{12}\right)\right]\,.
\end{align}

Given the renormalization group invariance of the hadronic tensor $\tilde M$ in impact parameter space,
\begin{align}
\tilde M = H(Q^2/\m^2)\, \tilde F_n(x;\vec b_\perp,Q,\m)\, \tilde F_\bn(z;\vec b_\perp,Q,\m)\ ,
\end{align}
we can establish the following relation between the anomalous dimension (AD) of the hard matching coefficient, $\g_H$, and the one of the TMDPDF, $\g_{n}$,
\begin{align}\label{eq:ads}
\g_n &= -\frac{1}{2} \g_H = -\frac{1}{2}A(\as)\ln\frac{Q^2}{\m^2} - \frac{1}{2} B(\as)\,.
\end{align}
Notice that this relation allows us to extract the AD of the TMDPDF up to third order in $\alpha_s$ from the known $\g_H$ at three loops~\cite{Moch:2005id,Moch:2004pa}.
And since $A(\as)=2\Gamma_{cusp}(\as)$ to all orders in perturbation theory, we get also
\begin{align}
\frac{d D(\as,L_T)}{d\ln\m} = \G_{cusp}(\as)\,.
\end{align}

Our master equation for the quark unpolarized TMDPDF  $\tilde F_{f/P}$ in impact parameter space is
\begin{align}
\tilde F_{f/P}\le(x,b;Q^2,\m=Q\ri) &=
%\stackrel{\rm A}{\overbrace{
\sum_{j=q,g} \int_x^1 \frac{dx'}{x'}\,
\tilde C^{\Qslash}_{f/j}\le(\frac{x}{x'},b;\m_{I}\ri)\,
\phi_{j/P}(x';\m_{I})
%}}
\nn\\
&\times
%\stackrel{\rm B}{\overbrace{
\exp \left\{
-D\le(b,\m_{I}\ri)\, \ln \frac{Q^2}{\mu_{I}^2}  +
\int_{\mu_{I}}^{\mu=Q} \frac{d \mu'}{\mu'} \left[ \g_n\le(\as(\m');\ln\frac{Q^2}{\m'^2}\ri) \right]\right\}
%}}
%\times
%\stackrel{\rm C}{\overbrace{
%\tilde F_{f/P}^{NP}\le(x,b;\m=Q,Q_0\ri)
%}}
\,,
\end{align}

%%%%%%%%%%%%%%%%%%%%%%%%%%%%%%%%
%%%%%%%%%%%%%%%%%%%%%%%%%%%%%%%%
%%%%%%%%%%%%%%%%%%%%%%%%%%%%%%%%
\section{Hard Part at ${\cal O}(\as)$}
\label{sec:hard}
%%%%%%%%%%%%%%%%%%%%%%%%%%%%%%%%
The hard matching coefficient for the $q_T$-dependent DY cross section is the same as the one for inclusive DY. As mentioned before, this matching coefficient at the higher scale $Q$ is obtained by matching the full QCD cross section onto the imaginary part of the product of two effective theory currents. This echoes the ``subtraction method'' in perturbative QCD.

We start by rewriting the cross section in a more useful way,
\begin{align} \label{eq:hadten}
d\sigma&=\frac{4 \pi\alpha}{3 N_c q^2 s}\frac{dx dz d^2 \vec q_\perp}{2 (2\pi)^4}
\sum_q e_q^2 M(x,z;\vec q_\perp,Q)\,,
\nn\\
M(x,z;\vec q_\perp,Q)&=
H(Q^2/\mu^2)
\int d^2\vec k_{n\perp} d^2\vec k_{\bn\perp}\,
\d^{(2)}(\vec q_\perp-\vec k_{n\perp}-\vec k_{\bn\perp})
\left[
%j_{n0} j_{\bn 0}
 \d(1-x)\d^{(2)}(\vec k_{n\perp})\d(1-z)\d^{(2)}(\vec k_{\bn\perp})
\right. \nn \\
&\left.
 + \a_s \left( F_{n1}\,\d(1-z)\d^{(2)}(\vec k_{\bn\perp}) +
  F_{\bn 1}\, \d(1-x)\d^{(2)}(\vec k_{n\perp}) \right)
\right] +O(\a_s^2)\,
\nn \\
&=H(Q^2/\mu^2)
\left[
\d(1-x)\d(1-z)\d^{(2)}(\vec q_\perp) \right.
\nn\\
&\left.
+
\a_s \Big(
\d(1-z)\, F_{n1}(x;\vec q_\perp,Q,\m) +
\d(1-x)\, F_{\bn 1}(z;\vec q_\perp,Q,\m)
\Big)
\right]
+O(\a_s^2)\,,
\end{align}
where $M$ is the hadronic tensor.

In QCD the virtual part of $M$ with the $\D$-regulator is
\begin{align}\label{eq:mqcdv}
M_{QCD}^v =
\frac{\a_s C_F }{2\pi}
\d(1-x)\d(1-z)\d^{(2)}(\vec q_\perp)
\left[ - 2\ln^2\frac{\D}{Q^2} - 3\ln\frac{\D}{Q^2} - \frac{9}{2} + \frac{\pi^2}{2} \right]\,.
\end{align}
The above result can be simply obtained by considering the one-loop correction to the vertex diagram for $q\bar q \to \g^*$, with the inclusion of the WFR diagram while using the fermion propagators in Eq.~(\ref{fermionsDelta}).

The virtual part of $F_n$ at one-loop is given in Eq.~(\ref{eq:jvn}), and we have an analogous result for $F_\bn$. Using Eq.~(\ref{eq:hadten}) the total virtual part of the hadronic tensor $M$ in the effective theory is
\begin{align}\label{eq:mscetv}
M_{SCET}^v &=
H(Q^2/\mu^2)
\frac{\a_s C_F}{2\pi}
\d(1-x)\d(1-z)\d^{(2)}(\vec q_\perp)
\left[
\frac{2}{\veuv^2} + \frac{1}{\veuv} \left( 3 + 2\ln\frac{\m^2}{Q^2}  \right) \right.
\nn\\
&
\left.
- 2\ln^2\frac{\D}{Q^2} - 3\ln\frac{\D}{Q^2}
+ 3\ln\frac{\m^2}{Q^2} + \ln^2\frac{\m^2}{Q^2} + \frac{7}{2} - \frac{2\pi^2}{3}
\right]\,,
\end{align}
where the UV divergences are canceled by the standard renormalization process. We notice that the IR contributions in Eqs.~(\ref{eq:mqcdv}) and~(\ref{eq:mscetv}) are the same, as they should, thus the matching coefficient between QCD and the effective theory at scale $Q$ is:
\begin{align}
H(Q^2/\mu^2) =
1 + \frac{\a_s C_F}{2\pi} \left[
- 3\ln\frac{\m^2}{Q^2} - \ln^2\frac{\m^2}{Q^2} - 8 + \frac{7\pi^2}{6}
\right]\,.
\end{align}
The above result was first derived in~\cite{Idilbi:2005ky}. We can also obtain the AD of the hard matching coefficient at ${\cal O}(\as)$ and verify Eq.~(\ref{eq:ads}),
\begin{align}
\g_{H1} = - \frac{\as C_F}{2\pi} \left[ 6 + 4\ln\frac{\m^2}{Q^2} \right] = -2\g_{n1}\,.
\end{align}

It is clear that the IR divergences of full QCD are recovered in the effective theory calculation, Eq.~(\ref{eq:mscetv}). To get this one must use the same regulator in both theories and the way that one regulates the IR physics (or more generally, the nUV) in both sides have to be consistent.
The matching coefficient at the higher scale depends only on the hard scale $Q^2$ as it should be.

%%%%%%%%%%%%%%%%%%%%%%%%%%%%%%%%
%%%%%%%%%%%%%%%%%%%%%%%%%%%%%%%%
%%%%%%%%%%%%%%%%%%%%%%%%%%%%%%%%%
\section{Gauge Invariance: $T$-Wilson Line}
%%%%%%%%%%%%%%%%%%%%%%%%%%%%%%%%%
One of the advantages of setting all the perturbative calculation  on-the-light cone is that  it is straightforward to show the gauge invariance of the relevant physical quantities.
Below we derive the transverse gauge link, the $T$ Wilson line, that has to be included in the effective theory in order to accomplish the calculation of collinear and soft matrix elements in a gauge invariant way~\cite{Idilbi:2010im,GarciaEchevarria:2011md}.

First we recall some of the features of the gluon fields in QCD in light-cone gauge (LCG)~\cite{Bassetto:1984dq}.
To fix matters, we work in QCD with the gauge fixing condition $\nb A=0$.
The canonical quantization of the gluon field proceeds by inserting in the Lagrangian the gauge fixing term
% \be \label{eq:l_gf}
$\mathcal{L}_{gf} = \Lambda^a (\nb A^a) $.
%\ee
 The $\Lambda^a$ is a field whose value on the Hilbert space of physical states is equal to zero.
  It is possible to write the most general solution of the equation of motion of the boson field $A_\mu^a$ by decomposing it as
  \begin{align} \label{eq:A_bassetto}
A^a_\m(k) &= T^a_\m(k)\d(k^2) + \nb_\m\frac{\d(\nb k)}{k_\perp^2} \L^a(n k,k_\perp)
%\nn \\ &
+
\frac{ik_\m}{k_\perp^2} \d(\nb k) U^a(n k,k_\perp)\,,
\end{align}
where the field $T^a_\m$ is such that $\nb^\m T^a_\m(k)=0$ and $k^\m T^a_\m(k)=0$.
 Fourier transforming this expression we see that in general the field $A^a_\m(x)$  has non-vanishing ``$-$''
and ``$\perp$'' (respectively $n A^{a}(x)$ and $ A^{\mu,a}_\perp(x)$) components when $x^-\rightarrow \pm \infty $.
Now we define
\begin{align}
 A^{(\infty)}  (x^+, x_\perp) &\stackrel{def}{=} A (x^+,\infty^-, x_\perp )\\
{\tilde A}(x^+,x^-, x_\perp) &\stackrel{def}{=} %\nn \\ &&
 A(x^+,x^-, x_\perp) -  A^{(\infty)} (x^+, x_\perp)
 \nn
  %=
%\nn\\
%{\tilde A}_{n &\stackrel{def}{=}  A_{n} -  A_{n}^{(\infty)}
\end{align}
and get
\be
i\Dslash_\perp = i\pdslash_\perp + g\Aslash_{\perp}=  i\pdslash_\perp +
g\tilde\Aslash_{\perp} + g\Aslash_{\perp}^{(\infty)}  \stackrel{def}{=}
i\tilde\Dslash_\perp  + g\Aslash_{\perp}^{(\infty)}\,.
\ee
Taking into account that the fields $\tilde A_\perp$ and  $A_{\perp}^{(\infty)}$ in Eq.~(\ref{dslash}) are evaluated at
 space-like separated points then one can show that
 \be \label{dslash}
i\Dslash_\perp= T i\tilde\Dslash_\perp T^\dagger\,,
\ee
 where
\begin{align}
\label{eq:Tqcd}
 T^\dagger%(x^+,x_\perp)=% \nn \\ &
 =P\exp\left[-i g\int_0^\infty d\tau { l}_\perp \cdot { A}_\perp^{(\infty)}(x^+,{ x}_\perp-{l}_\perp\tau)
\right]\,,
 \end{align}

This equation  leads us automatically to include the
$T$-Wilson line at the level of the soft-collinear effective theory (SCET) Lagrangian.
Moreover
$1/\bn\pd$ and $T$ commute because the $T$-Wilson line does not depend on $x^-$. Under gauge transformation
$\delta A_{\mu\perp}^{(\infty)}=D_{\mu\perp}\omega$   one has
$T(x^+, x_\perp)\rightarrow U(x^+, x_\perp) T(x^+, x_\perp) U^\dagger(x^+, x_\perp-l_\perp\infty)= U(x^+, x_\perp) T(x^+, x_\perp)$
 since $A_{\mu\perp}^{(\infty)}(x^+, \infty_\perp)=0 $. Notice also that the $T$-Wilson lines are  independent of $l_\perp$.

Now we split the fermion field into large and small components using the usual projectors $\nslash \bnslash/4$  and $\bnslash \nslash/4$ and eliminate the small components using the equations of motion~\cite{Bauer:2000yr}.
The result of this is
\begin{align}
\label{eq:LLCG1}
\mathcal{L} &= \bar \xi_n \left( inD + i\Dslash_\perp \frac{1}{i\bn D}
i\Dslash_\perp \right) \frac{\bnslash}{2}\xi_n\ . %\nn \\ &
\end{align}
In QCD and in LCG with the gauge condition $\bn A=0$ we get
\begin{align}
\label{eq:LLCG2}
\mathcal{L} &=
\bar \xi_n \left( inD + T i\tilde\Dslash_\perp
\frac{1}{i\bn \pd} i\tilde\Dslash_\perp T^\dagger \right)\frac{\bnslash}{2} \xi_n\,.
\end{align}
 In order to get the SCET Lagrangian we must implement multipole expansion and power counting on the fields that appear
in Eq.~(\ref{eq:LLCG1}). In SCET we have also the freedom to choose a different gauge in the different sectors of the theory.
 We distinguish the cases of SCET-I and SCET-II.  The two formulations differ essentially in the scaling of the soft sector of the
 theory.
Here  we are interested in SCET-II which is the effective theory necessary for the Drell-Yan $q_T$-spectrum where $q_T$ is much larger than $\Lambda_{\rm QCD}$ and much smaller than $Q$ .
The collinear modes we have in In SCET-II scale as $(\nb k, n k, k_\perp)\sim Q(1,\eta^2 ,\eta)$ where $\eta\ll 1$ and the soft modes as
$(\nb k, n k, k_\perp)\sim Q(\eta,\eta,\eta)$.
Collinear and soft gluon fields' components scale accordingly.

%%%%%%%%%%%%%%%%%%%%%%%%%%%%%%%%%%%%%%%%%%%%%%%%%%%%%%%%%%%%%%%%%%%%%%%%%%%%%%%%%%%%%%%%%%%%%%%%%%%%%%%%%%%
%
%%%%%%%%%%%%%%%%%%%%%%%%%%%%%%%%%%%%%%%%%%%%%%%%%%%%%%%%%%%%%%%%%%%%%%%%%%%%%%%%%%%%%%%%%%%%%

%\underline{\textit {\bf SCET-II.}}
%The analysis of the collinear  sector in SCET-II is the same as for SCET-I.
The soft sector  of SCET-II  has peculiar features. In regular gauges soft partons  do not interact with collinear  ones because
the interactions knock the collinear fields far off-shell. This is also true in light-cone gauge except when
 one makes the choice $n A_s=0$ (take  here a covariant gauge for collinear fields for fixing ideas).
 It is  easy  to be convinced that interactions like $\prod_i \phi_n^i(x)A_{s\perp}^\infty (x^-,x^\perp)$, where here ``$\infty$'' refers to the $+$ direction and
$\phi_n^i(x)$ are generic collinear fields, preserve the on-shellness of the collinear particles.
Using multipole expansion the  vertex  becomes $\prod_i \phi_n^i(x)A_{s\perp}^\infty (0^-,x^\perp)$
(because for collinear fields $x^-\sim 1$ and for the soft field $x^-\sim 1/\eta$).
In this gauge the covariant derivative for collinear particles  becomes
 $i D^\mu=i\partial^\mu+g A^\mu_n(x)+g A_{s\perp}^{(\infty)\mu}(0^-,x_\perp)$. The gauge ghost $A_{s\perp}^{(\infty)}$ however can be decoupled
 from collinear gluons defining a ``soft free'' collinear gluon $A^{(0)\mu}_n(x)= T_{s n} (x_\perp)A^\mu_n(x)T_{s n}^\dagger(x_\perp)$ where
\begin{align}
\label{eq:Tsoft}
 T_{s n} %(x^+,x_\perp)=% \nn \\ &
 =\bar P\exp\left[i g\int_0^\infty d\tau { l}_\perp \cdot { A_s}_\perp^{(\infty)}(0^-,{ x}_\perp-{ l}_\perp\tau)
\right]\, .
\end{align}
Defining $D^{(0)\mu}_n=i\partial^\mu+g A^{(0)\mu}_n $  we  have  $il_\perp D_{\perp}=T_{s n} (x_\perp)il_\perp D_{n\perp}^{(0)}T_{s n}^\dagger (x_\perp) $ and
\begin{align}
\label{eq:SCETIii}
\mathcal{L} &=
\bar \xi_{n}^{(0)} \Big( in D_{ n}^{(0)}
%\right. \nn \\
%&\left.
 +i\Dslash_{ n\perp}^{(0)}W_{n}^{T(0)}
\frac{1}{i\bn \partial}  W_{n}^{T(0)\dagger} i\Dslash_{ n\perp}^{(0)}  \Big)\frac{\bnslash}{2} \xi_{ n}^{(0)},
\end{align}
where $\xi_{ n}^{(0)}= T_{s n} (x_\perp) \xi_n(x)$ and $ W_{n}^{T(0)}=T_n^{(0)} W^{(0)}$ are made out of soft free gluons.
Thus, and thanks to $T_{s n}$ Wilson lines, the soft
partons are completely decoupled from the collinear ones.

%\begin{figure}
%\begin{center}
%\includegraphics[width=0.7\textwidth]{s_virtuals_lc}
%\end{center}
%\caption{\it The  soft  function at one-loop in light-cone gauge.
%\label{s_virtuals_lc}}
%\end{figure}

With the existence of the $T$-Wilson lines it is possible now to compare the one-loop contributions in both Feynman gauge and LCG. This comparison can be performed at the level of the integrands without the need to perform any actual calculations.

%In this Section we show that the TMDPDFs, $J_{n(\bn)}$, are actually  the same in light-cone  gauge and Feynman gauge, once the contribution from the transverse Wilson lines is taken into %account. In Ref.~\cite{Idilbi:2010im} two of us have shown that the naive collinear contribution to the TMDPDF (the numerator in $j_{n(\bn)}$) is actually  gauge invariant
% with a one-loop calculation.  In that article  the authors used a particular IR regulator for light-cone divergences
%however the results obtained in covariant gauge and in light-cone gauge are the same  and independent of that regulator once the zero-bin corrections are included. That was shown explicitly
%in the Appendix of that work.

 In light-cone gauge we use the ML prescription~\cite{Mandelstam:1982cb}, which is the only one consistent
with the canonical quantization of  QCD in this gauge~\cite{Bassetto:1984dq}.
 Moreover in the $n$ and $\bn$ collinear sectors the only gauge fixings compatible with
the power counting of the collinear particles are respectively $\bn A_n=0$ and $n A_\bn=0$,
which correspond to ``killing'' the highly oscillating component of the gluon field in each sector.
We now compare  the integrals  that we have evaluated in Feynman gauge with the  corresponding ones in light-cone gauge.

The interesting contribution to the collinear part of the TMDPDF in Feynman gauge is provided  by
the $W_n$ Wilson line and it is  (cfr.~Eq.~(\ref{eq:jn1c}))
\begin{align}
\label{eq:vFey}
\hat f_{n1}^{(\ref{n_virtuals}c)\; (Feyn)}=
-\d(1-x)\d^{(2)}(\vec k_{n\perp}) 2i g^2 C_F \mu^{2\eps}
\int  \frac{d^d k}{(2\pi)^d}\; \frac{1}{(k^2+i0)( k^+-i0)}\frac{ p^++ k^+}{(p+k)^2+i0}\ .
\end{align}
In light-cone gauge this result  is reproduced  when one combines  the axial part of the WFR
\begin{align}
\label{eq:IAx}
\hat f_{n1}^{(\ref{n_virtuals}a)\; (Ax)}=
\d(1-x)\d^{(2)}(\vec k_{n\perp}) 4 ig^2 C_F \mu^{2\eps}
\int \frac{d^d k}{(2\pi)^d}\; \frac{1}{(k^2+i0)}\frac{ p^++ k^+}{(p+k)^2+i0}\Big[\frac{\theta(k^-)}{k^++i0}+\frac{\theta(-k^-)}{k^+-i0}\Big]\ ,
\end{align}
with the contribution of the $T$ Wilson line
\begin{align}
\label{eq:IT}
\hat f_{n1}^{(\ref{n_virtuals}c)\; (T)}=
- \d(1-x)\d^{(2)}(\vec k_{n\perp}) 2 ig^2 C_F \mu^{2\eps}
\int \frac{d^d k}{(2\pi)^d}\; \frac{1}{(k^2+i0)}\frac{ p^++ k^+}{(p+k)^2+i0}\theta(k^-)\Big[\frac{1}{k^+-i0}-\frac{1}{k^++i0}\Big]\ .
 \end{align}
It is clear that $\hat f_{n1}^{(\ref{n_virtuals}c)\; (Feyn)}=\hat f_{n1}^{(\ref{n_virtuals}c)\; (T)}-\hat f_{n1}^{(\ref{n_virtuals}a)\; (Ax)}/2$.
In SCET there is also a tadpole diagram that needs to be considered however it is null in light cone gauge since the gluon field does not propagate at infinity~\cite{Idilbi:2010im}.
Similar considerations hold for Feynman diagrams with real gluon contributions and for the soft sector~\cite{GarciaEchevarria:2011rb}.

As a final remark, it is not so obvious how one can implement light-cone gauge calculations while going off-the-light-cone.

%%%%%%%%%%%%%%%%%%%%%%%%%%%%%%%%%
%%%%%%%%%%%%%%%%%%%%%%%%%%%%%%%%%
\section{Conclusions}
%%%%%%%%%%%%%%%%%%%%%%%%%%%%%%%%%
We have argued that to properly define the TMDPDF, one needs to include the soft factor (a square root of it). With this inclusion the TMDPDF can be renormalized and it captures all the IR of full QCD when it is introduced in a properly obtained factorization theorem.
We argued that one does not need to introduce any of the ``off-the-light-cone'' contributions, while still being able to resum large logarithms, thus avoiding the use of the Collins-Soper evolution equation.

The evolution of the TMDPDF has been discussed and the $Q^2$-resummation at the intermediate scale was explained. That was achieved under the assumption of identical treatment of two collinear sectors and it was mainly based on the fact that the TMDPDF is free from rapidity divergences.

The conclusions above are independent of the set of regulators that one uses for the nUV. Moreover, if one is consistent with the regularization of nUV divergencies in the theories above and below a certain relevant scale, then the matching coefficients will be independent of this set of regulators.

We also discussed the relevance of the $T$-Wilson line and their contribution to render hadronic matrix elements gauge invariant. Contrary to many existing statements in the literature, those Wilson lines are crucial for the gauge invariance to hold.
Finally, all this effort can be extended to the proper definition and treatment of polarized quark and gluon TMDs, which are important ingredients in order to get information about the inner structure of the nucleons.

%%%%%%%%%%%%%%%%%%%%%%%%%%%%%%%%%
%%%%%%%%%%%%%%%%%%%%%%%%%%%%%%%%%
\section*{Acknowdlegements}
%%%%%%%%%%%%%%%%%%%%%%%%%%%%%%%%%
This work is supported by the Spanish MEC, FPA2011-27853-CO2-02.
M.G.E. is supported by the PhD funding program of the Basque Country Government.
A.I. is supported by BMBF (06RY9191).
I.S. is supported by the Ram\'on y Cajal Program.

%%%%%%%%%%%%%%%%%%%%%%%%%%%%%%%%%
%%%%%%%%%%%%%%%%%%%%%%%%%%%%%%%%%


\begin{thebibliography}{99}
%%%%%%%%%%%%%%%%%%%%%%%%%%%%%%%%%

%\cite{Barone:2010zz}
\bibitem{Barone:2010zz}
  V.~Barone, F.~Bradamante and A.~Martin,
  %``Transverse-spin and transverse-momentum effects in high-energy processes,''
  Prog.\ Part.\ Nucl.\ Phys.\  {\bf 65} (2010) 267
  [arXiv:1011.0909 [hep-ph]].
  %%CITATION = ARXIV:1011.0909;%%

  %\cite{Manohar:2006nz}
\bibitem{Manohar:2006nz}
  A.~V.~Manohar and I.~W.~Stewart,
  %``The Zero-Bin and Mode Factorization in Quantum Field Theory,''
  Phys.\ Rev.\ D {\bf 76} (2007) 074002
  [hep-ph/0605001].
  %%CITATION = HEP-PH/0605001;%%

  %\cite{Chay:2012mh}
\bibitem{Chay:2012mh}
  J.~Chay and C.~Kim,
  %``Structure of divergences in Drell-Yan process with small transverse momentum,''
  arXiv:1208.0662 [hep-ph].
  %%CITATION = ARXIV:1208.0662;%%

   %\cite{Collins:2011zzd}
\bibitem{Collins:2011zzd}
  J.~Collins,
  ``Foundations of perturbative QCD,''
  (Cambridge monographs on particle physics, nuclear physics and cosmology. 32)

%\cite{GarciaEchevarria:2011rb}
\bibitem{GarciaEchevarria:2011rb}
  M.~G.~Echevarria, A.~Idilbi and I.~Scimemi,
  %``Factorization Theorem For Drell-Yan At Low q_T And Transverse Momentum Distributions On-The-Light-Cone,''
  JHEP {\bf 1207} (2012) 002
  [arXiv:1111.4996 [hep-ph]].
  %%CITATION = ARXIV:1111.4996;%%

%\cite{Echevarria:2012pw}
\bibitem{Echevarria:2012pw}
  M.~G.~Echevarria, A.~Idilbi, A.~Sch\"afer and I.~Scimemi,
  %``Model-Independent Evolution of Transverse Momentum Dependent Distribution Functions (TMDs) at NNLL,''
  arXiv:1208.1281 [hep-ph].
  %%CITATION = ARXIV:1208.1281;%%
  
  %\cite{Lee:2006nr}
\bibitem{Lee:2006nr}
  C.~Lee and G.~F.~Sterman,
  %``Momentum Flow Correlations from Event Shapes: Factorized Soft Gluons and Soft-Collinear Effective Theory,''
  Phys.\ Rev.\ D {\bf 75} (2007) 014022
  [hep-ph/0611061].
  %%CITATION = HEP-PH/0611061;%%
  
  %\cite{Idilbi:2007ff}
\bibitem{Idilbi:2007ff}
  A.~Idilbi and T.~Mehen,
  %``On the equivalence of soft and zero-bin subtractions,''
  Phys.\ Rev.\ D {\bf 75} (2007) 114017
  [hep-ph/0702022 [HEP-PH]].
  %%CITATION = HEP-PH/0702022;%%
  
  %\cite{Idilbi:2007yi}
\bibitem{Idilbi:2007yi}
  A.~Idilbi and T.~Mehen,
  %``Demonstration of the equivalence of soft and zero-bin subtractions,''
  Phys.\ Rev.\ D {\bf 76} (2007) 094015
  [arXiv:0707.1101 [hep-ph]].
  %%CITATION = ARXIV:0707.1101;%%

   %\cite{Collins:1984kg}
\bibitem{Collins:1984kg}
  J.~C.~Collins, D.~E.~Soper and G.~F.~Sterman,
  %``Transverse Momentum Distribution in Drell-Yan Pair and W and Z Boson Production,''
  Nucl.\ Phys.\ B {\bf 250} (1985) 199.
  %%CITATION = NUPHA,B250,199;%%

\bibitem{Becher:2010tm}
  T.~Becher and M.~Neubert,
  %``Drell-Yan production at small q_T, transverse parton distributions and the collinear anomaly,''
  Eur.\ Phys.\ J.\ C {\bf 71}, 1665 (2011)
  [arXiv:1007.4005 [hep-ph]].
  %%CITATION = ARXIV:1007.4005;%%

%\cite{Aybat:2011zv}
\bibitem{Aybat:2011zv}
  S.~M.~Aybat and T.~C.~Rogers,
 % ``TMD Parton Distribution and Fragmentation Functions with QCD Evolution,''
  Phys.\ Rev.\ D {\bf 83} (2011) 114042
  [arXiv:1101.5057 [hep-ph]].
  %%CITATION = ARXIV:1101.5057;%%

%\cite{Aybat:2011ge}
\bibitem{Aybat:2011ge}
  S.~M.~Aybat, J.~C.~Collins, J.~-W.~Qiu and T.~C.~Rogers,
  %``The QCD Evolution of the Sivers Function,''
  Phys.\ Rev.\ D {\bf 85} (2012) 034043
  [arXiv:1110.6428 [hep-ph]].
  %%CITATION = ARXIV:1110.6428;%%

%\cite{Aybat:2011ta}
\bibitem{Aybat:2011ta}
  S.~M.~Aybat, A.~Prokudin and T.~C.~Rogers,
  %``Calculation of TMD Evolution for Transverse Single Spin Asymmetry Measurements,''
  arXiv:1112.4423 [hep-ph].
  %%CITATION = ARXIV:1112.4423;%%

%\cite{Anselmino:2012aa}
\bibitem{Anselmino:2012aa}
  M.~Anselmino, M.~Boglione and S.~Melis,
  %``A Strategy towards the extraction of the Sivers function with TMD evolution,''
  arXiv:1204.1239 [hep-ph].
  %%CITATION = ARXIV:1204.1239;%%


%\cite{Ji:2004wu}
\bibitem{Ji:2004wu}
  X.~-d.~Ji, J.~-p.~Ma and F.~Yuan,
  %``QCD factorization for semi-inclusive deep-inelastic scattering at low transverse momentum,''
  Phys.\ Rev.\ D {\bf 71} (2005) 034005
  [hep-ph/0404183].
  %%CITATION = HEP-PH/0404183;%%

  %\cite{Sun:2011iw}
\bibitem{Sun:2011iw}
  P.~Sun, B.~-W.~Xiao and F.~Yuan,
  %``Gluon Distribution Functions and Higgs Boson Production at Moderate Transverse Momentum,''
  Phys.\ Rev.\ D {\bf 84} (2011) 094005
 [arXiv:1109.1354 [hep-ph]].
  %%CITATION = ARXIV:1109.1354;%%

  %\cite{Moch:2005id}
\bibitem{Moch:2005id}
  S.~Moch, J.~A.~M.~Vermaseren and A.~Vogt,
  %``The Quark form-factor at higher orders,''
  JHEP {\bf 0508} (2005) 049
  [hep-ph/0507039].
  %%CITATION = HEP-PH/0507039;%%

  %\cite{Moch:2004pa}
\bibitem{Moch:2004pa}
  S.~Moch, J.~A.~M.~Vermaseren and A.~Vogt,
  %``The Three loop splitting functions in QCD: The Nonsinglet case,''
  Nucl.\ Phys.\ B {\bf 688} (2004) 101
  [hep-ph/0403192].
  %%CITATION = HEP-PH/0403192;%%

  %\cite{Idilbi:2005ky}
\bibitem{Idilbi:2005ky}
  A.~Idilbi and X.~-d.~Ji,
  %``Threshold resummation for Drell-Yan process in soft-collinear effective theory,''
  Phys.\ Rev.\ D {\bf 72} (2005) 054016
  [hep-ph/0501006].
  %%CITATION = HEP-PH/0501006;%%

%\cite{Idilbi:2010im}
\bibitem{Idilbi:2010im}
  A.~Idilbi and I.~Scimemi,
  %``Singular and Regular Gauges in Soft Collinear Effective Theory: The Introduction of the New Wilson Line T,''
  Phys.\ Lett.\ B {\bf 695} (2011) 463
  [arXiv:1009.2776 [hep-ph]].
  %%CITATION = ARXIV:1009.2776;%%

  %\cite{GarciaEchevarria:2011md}
\bibitem{GarciaEchevarria:2011md}
  M.~Garcia-Echevarria, A.~Idilbi and I.~Scimemi,
  %``SCET, Light-Cone Gauge and the T-Wilson Lines,''
  Phys.\ Rev.\ D {\bf 84} (2011) 011502
 [arXiv:1104.0686 [hep-ph]].
  %%CITATION = ARXIV:1104.0686;%%

%\cite{Mandelstam:1982cb}
\bibitem{Mandelstam:1982cb}
  S.~Mandelstam,
  %``Light Cone Superspace and the Ultraviolet Finiteness of the N=4 Model,''
  Nucl.\ Phys.\ B {\bf 213} (1983) 149.
  %%CITATION = NUPHA,B213,149;%%

  %\cite{Bassetto:1984dq}
\bibitem{Bassetto:1984dq}
  A.~Bassetto, M.~Dalbosco, I.~Lazzizzera and R.~Soldati,
  %``Yang-Mills Theories in the Light Cone Gauge,''
  Phys.\ Rev.\ D {\bf 31} (1985) 2012.
  %%CITATION = PHRVA,D31,2012;%%

  %\cite{Bauer:2000yr}
\bibitem{Bauer:2000yr}
  C.~W.~Bauer, S.~Fleming, D.~Pirjol and I.~W.~Stewart,
  %``An Effective field theory for collinear and soft gluons: Heavy to light decays,''
  Phys.\ Rev.\ D {\bf 63} (2001) 114020
  [hep-ph/0011336].
  %%CITATION = HEP-PH/0011336;%%


\end{thebibliography}
\end{document}